\documentclass[printformat]{aa}
\usepackage{amsmath}
\usepackage{graphicx}
\usepackage{times}

\usepackage{graphicx,txfonts,natbib}
\usepackage{longtable}
\usepackage{lscape}

\newcommand{\cd}{d\,$^{-1}$\,}

\begin{document}
\title{Seismic modelling of the $\beta\,$Cep star HD\,180642 (V1449\,Aql)} 

\author{
C.~Aerts\inst{1,2}
\and M.~Briquet\inst{1,3,}\thanks{Postdoctoral Fellow of the Fund for Scientific
Research, Flanders} 
\and P.~Degroote\inst{1}
\and A.~Thoul\inst{4,}\thanks{Chercheur Qualifi\'e, FRS-F.N.R.S., Belgium}
\and T.~Van Hoolst\inst{1,5}
\offprints{C.\ Aerts}
}

\institute{
Instituut voor Sterrenkunde, K.U.Leuven, Celestijnenlaan 200D, B-3001
Leuven, Belgium 
\and Department of Astrophysics, IMAPP, Radboud University Nijmegen, 
PO Box 9010, 6500 GL Nijmegen, The Netherlands
\and
LESIA, Observatoire de Paris, CNRS, UPMC, Universit\'e Paris-Diderot, 92195
Meudon, France 
\and
Institut d'Astrophysique et de G\'eophysique, Universit\'e de Li\`ege, All\'ee
du 6 Ao\^ut 17, B-4000 Li\`ege, Belgium
\and
Royal Observatory of Belgium, Ringlaan 3, B-1180 Brussels, Belgium}

\date{Received ; accepted}
\authorrunning{Aerts et al.}
\titlerunning{Seismic modelling of the $\beta\,$Cep star HD\,180642}

\abstract{We present modelling of the $\beta\,$Cep star HD\,180642
  based on its observational properties deduced from CoRoT and ground-based
  photometry as well as from time-resolved spectroscopy.}  {We
  investigate whether present-day state-of-the-art 
models are able to explain
  the full seismic behaviour of this star, which has extended observational
  constraints for this type of pulsator.}  {We constructed a dedicated database
  of stellar models and their oscillation modes tuned to fit the dominant radial
  mode frequency of HD\,180642, by means of 
varying the hydrogen content, metallicity, mass,
  age,
  and core overshooting parameter. We compared the seismic properties of these
  models with those observed.}  {We find models that are able to
  explain the numerous 
observed oscillation properties of the star, for a narrow range in
  mass of 11.4--11.8\,M$_\odot$ and no or very mild overshooting (with up to
  0.05 local pressure scale heights), except for an excitation problem of the
  $\ell=3$, p$_1$ mode. We deduce a rotation period of about 13\,d, which is
  fully compatible with recent magnetic field measurements.  The seismic models
  do not support the earlier claim of solar-like oscillations in the
  star. We instead ascribe the power excess at high frequency to non-linear
  resonant mode coupling between the high-amplitude radial fundamental mode and
  several of the low-order pressure modes.  We report a discrepancy between the
  seismic and spectroscopic gravity at the $2.5\sigma$ level.}{}
\maketitle

%

\section{Introduction}

On the basis of observations 
by the small satellites WIRE (Buzasi 2004) and MOST (Matthews 2007),
high-precision time-resolved space photometry has been obtained by the ongoing
missions CoRoT (Baglin et al.\ 2006; Auvergne et al.\ 2009) and {\it Kepler\/}
(Burocki et al.\ 2010; Gilliland et al.\ 2010) providing significant progress in
particular subfields of asteroseismology. Notable discoveries have been made
of oscillation
properties unknown prior to the era of space asteroseismology for
solar-like oscillations, not only in both FGK-type main-sequence stars
(e.g., Michel et al.\ 2008, Chaplin et al.\ 2011) and red giants (e.g., De
Ridder et al.\ 2009, Beck et al.\ 2011, Bedding et al.\ 2011), but also in a
massive O-type star (Degroote et al.\ 2010b).

Thanks to the continuity between space- and ground-based observations, 
the
discovery of hybrid pulsators in various evolutionary stages, i.e., stars
pulsating in pressure (p) and gravity (g) modes simultaneously, will also shed
new light on the details of stellar structure models. While a few of these hybrid
pulsators had been known prior to the space missions, large amounts of them
were found from the space photometry, all along the main sequence from spectral
type B (e.g., Balona et al.\ 2011) to F (Grigahc\`ene et al.\ 2010).  
Seismic diagnostics suitable for interpreting these cases in terms of modes
probing both the core region and the outer envelope are yet to be developed, but
it is clear that the numerous gravity modes offer a unique opportunity to test
the physics in the deep interior regions adjacent to the core. Initial efforts to
explore this region concluded that the input physics of the current models
has to be improved to accurately describe 
chemically inhomogeneous regions in order to
be compliant with the space data (Degroote et al.\ 2010a).

For the most massive pulsators of spectral types O and {B0--2}, space
photometry so far provided more questions than answers. While the detection of
pulsations in photometric data was finally established for O-type stars, there
seems to be a different cause of the pulsations. The O9V star HD\,46202 behaves
in a similar way to 
known $\beta\,$Cep stars (e.g., Aerts et al.\ 2010 for an extensive
overview of all types of pulsators across the Hertzsprung-Russell diagram) but
none of its modes is predicted to be excited by the theory that is able to 
explain the $\beta\,$Cep instability strip (Briquet et al.\
2011). Moreover, the O8.5V star HD\,46149 appears to oscillate in 
stochastic modes
fulfilling scaling laws and \'echelle diagrams with clear ridges for modes of
equal degree but consecutive radial orders 
as those observed for Sun-like
stars (Degroote et al.\ 2010b). This implies that this star must have an
outer convection zone of considerable size, 
a suggestion made for massive stars in
general, independently of seismic data, from theoretical work (Cantiello et al.\
2009). Three hotter O stars do not show any coherent classical pulsational
signal, but rather an excess power at low frequency resembling red noise (Blomme
et al.\ 2011). More accurate knowledge of the physics in the outer envelope
of O stars is needed to understand the variability behaviour detected from
CoRoT (there are no O stars in the {\it Kepler\/} FoV).

Few B0--2 pulsators with or without emission lines have been 
observed by CoRoT, while most of the  
B/Be stars monitored by {\it Kepler\/} are cooler.
A remarkable finding is the outburst signal detected in the B0.5IVe star
HD\,49330 (Huat et al.\ 2009), a phenomenon that has not been observed to 
occur in any of the
other observed Be stars, which are of later spectral type.  Theoretical
work is needed to understand and interpret the observed frequency patterns in
the Be pulsators (Neiner et al.\ 2009, Diago et al.\ 2009, Guti\'errez-Soto et
al.\ 2009).  Progress is hampered by the mathematical complexity of describing 
the
oscillations in these flattened stars, induced by their
rapid rotation, often close to their critical velocity.

To date CoRoT has 
observed two slow rotators without emission lines situated in the
$\beta\,$Cep instability strip. The star B0.5V star HD\,51756 turned out to be a
rotational modulation variable rather than a pulsator (P\'apics et al.\ 2011).
In this work, we present seismic modelling of the only known $\beta\,$Cep class
member monitored by CoRoT: the B1.5II-III star HD\,180642 (V1449\,Aql).

\section{Summary of observational constraints}

\begin{table}
\caption{Frequencies and mode identification of the highest amplitude modes of
    HD\,180642.}
\label{freqs}
\centering
\begin{tabular}{ccccrr}
\hline
Frequency & MI method & $\ell$ & $m$ & $i$ & $V_{\rm eq}$ \\
(\cd) & & & & $(^\circ)$ & (km\,s$^{-1}$) \\
\hline
5.48694 & ampl.ratios & 0 & 0  & & \\
7.35867 & ampl.ratios & 0 or 3 &&\\
8.40790 & LPVs & 3 & $\pm$2 & 82 &38\\
&\ \ \ \ \ \ \ \ \ or& 3 & $\pm$1 & 31 &78\\
&\ \ \ \ \ \ \ \ \ or& 1 & $\pm$1 & 15 &119\\
&\ \ \ \ \ \ \ \ \ or& 2 & $\pm$1 & 14 &104\\
&\ \ \ \ \ \ \ \ \ or& 2 & $\pm$1 & 76 &26\\
&\ \ \ \ \ \ \ \ \ or& 2 & $\pm$2 & 48 &52\\
&\ \ \ \ \ \ \ \ \ or& 3 & $\pm$3 & 67 &26\\
6.14336 &none&\ldots&\ldots&\ldots&\ldots\\
6.26527 &none&\ldots&\ldots&\ldots&\ldots\\
6.32482 &none&\ldots&\ldots&\ldots&\ldots\\
7.10353 &none&\ldots&\ldots&\ldots&\ldots\\
7.25476 &none&\ldots&\ldots&\ldots&\ldots\\
8.77086 &none&\ldots&\ldots&\ldots&\ldots\\
\hline
0.29917 &none&\ldots&\ldots&\ldots&\ldots\\
0.89870 &none&\ldots&\ldots&\ldots&\ldots\\
\hline
\end{tabular}\tablefoot{MI method can be any of: photometric 
amplitude ratios (ampl.rations), line-profile variations (LPVs), not available
(none). }
\end{table}

The $\beta\,$Cep class member HD\,180642 has a visual magnitude of 8.3 and a
dominant radial mode with $V$ amplitude 39\,mmag (Aerts 2000).  The additional
observational characteristics of HD\,180642 we used as input for the seismic
modelling were taken from Degroote et al.\ (2009, Paper\,I) and Briquet et al.\
(2009, Paper\,II).  Paper\,I was devoted to the analysis of the CoRoT light
curve of the star. Three versions of a model description for the light curve
were considered: one with classical prewhitening producing 127 significant
frequencies, one with harmonics and combination frequencies of 11 independent
frequencies, and one where time-dependent amplitudes and phases were allowed.
After careful statistical evaluation taking into account 
penalties for the number of free
parameters, the best fit to the CoRoT light curve was found to be
the model based on
33 frequencies among which 11 were independent ones (which are repeated in
Table\,\ref{freqs}), 3 were 
harmonics of the dominant frequency and 19 were combination
frequencies (listed in Table\,2 of Paper\,I and not repeated here).

The list of 
nine highest-amplitude independent stable frequencies contains the one of the
dominant radial mode known prior to the CoRoT launch (Aerts 2000), eight
frequencies in the range $[6.1,8.5]\,$\cd, and 
two frequencies below 1\cd, which could
either be due to g modes or connected to the rotation of the star (or both)
and agree to within a factor three.

These ten frequencies have amplitudes roughly between 0.8 and 6\,mmag (Paper\,I),
i.e., far below the 39\,mmag of the dominant mode.  The frequency spectrum was
found to be atypical of a $\beta\,$Cep star, in the sense that ten sum and
nine difference frequencies were also detected, in addition to three 
harmonics of
the dominant frequency.  Moreover, several of the combination frequencies were
found to have locked phases, a characteristic expected for modes undergoing 
non-linear mode interaction with the large-amplitude dominant oscillation.
The 11 independent largest-amplitude frequencies found in the space photometry
are listed in Table\,\ref{freqs}, along with their 
mode identification obtained by Aerts (2000) and in Papers\,I and II. 
In this work, we consider the nine highest
frequencies in that table for the modelling procedure, i.e., 
the  two frequencies below 1\,\cd are not used.
The total range of frequencies related 
to dominant independent modes and their combinations covers
$[0.3,27.5]\,$\cd.

In addition to these 33 frequencies, variations with time-dependent behaviour
were found in the residual light curve, within the frequency range
of the harmonics and the combination frequencies of the heat-driven modes
(Belkacem et al.\ 2009; Paper\,I).  Belkacem et al.\ (2009) determined a
frequency spacing of 1.17\,\cd for the range above 11.23\,d$^{-1}$ and below
26\,\cd and interpreted it in terms of solar-like oscillations caused by
stochastic forcing due to convective motions in the outer stellar layers.  In
Paper\,I, a value for candidate frequency spacings was recomputed over the range
$[4.32,25.9]\,$\cd for the three types of prewhitening procedures and led to
1.05, 1.11, and 1.56\,\cd, with an uncertainty of $\sim\,0.02\,$\cd (the
corresponding four spacing values are 13.5, 12.1, 12.9, and 18.0\,$\mu$Hz).  The
physical interpretation of these spacings remains unclear, given that their
value depends on the way the CoRoT light curve is prewhitened. We return to
this problem in Sect.\,\ref{spacing}.

In Paper\,II, 
we presented the results of 
an extensive ground-based multicolour photometric and high-resolution
spectroscopic campaign that we have organised.
This campaign has led 
to the detection of three of the CoRoT frequencies in the photometry and nine in
the spectroscopy, among then the fourth harmonic of the dominant frequency, which
is not found in the space photometry. Empirical mode identifation, based on
photometric 
amplitude ratios and on line-profile variability, led to the results in
Table\,\ref{freqs}. In particular, the frequency 7.35867\,\cd was identified 
with either an $\ell=0$ or $\ell=3$ mode from amplitude ratios, while the moment
method was used to deduce information on the mode wavenumbers $(\ell,m)$, the
inclination angle, and the equatorial velocity for the frequency
8.40790\,\cd. The seven 
possibilities listed in Table\,\ref{freqs} occurred with a
much higher probability than any of the other options for $(\ell,m)$ and we
allow any of those seven solutions in the modelling rather than taking only the
best one.

The average spectrum of HD\,180642 was also analysed in Paper\,II to determine
the fundamental parameters of the star, as well as its current surface 
abundance pattern. This
resulted in $T_{\rm eff} = 24\,500 \pm 1\,000$K, $\log\,g=3.45 \pm 0.15$, and
$Z\in
[0.0083,0.0142]$, where the interval for $Z$ covers the dependence on the
microturbulent velocity that could not be easily fixed and led to a systematic
uncertainty.

We also recall from Paper\,II that the overall line broadening amounts
  to 44\,km\,s$^{-1}$. This is the combined effect of line broadening caused by 
the
  projected rotation velocity and the pulsations. It is thus an upper limit to
  $v\sin i$.  The dominant mode has a radial-velocity amplitude of
  39\,km\,s$^{-1}$ (Paper\,II), which, taking into account limb-darkening
  effects,  translates the overall broadening value into an
  upper limit of $\sim\,30$\,km\,$s^{-1}$ for $v\sin i$.

  Finally, we point out that Hubrig et al.\ (2011) obtained and analysed
  spectropolarimetric time series observations taken with the SOFIN \'echelle
  spectrograph at the 2.56m Nordic Optical Telescope, after the discovery of a
  magnetic field in the star (Hubrig et al.\ 2009). From their 13 data points,
  three dominant peaks emerged in the periodogram, with the highest peak at
  0.072\,\cd. This periodicity was interpreted in the framework of a rigid
  rotator model for which the period of the magnetic field variation corresponds
  to a stellar rotation period of 13.9\,d, assuming a centred dipole.
\begin{figure*}[t!]
\begin{center}
\rotatebox{0}{\resizebox{9cm}{!}{\includegraphics{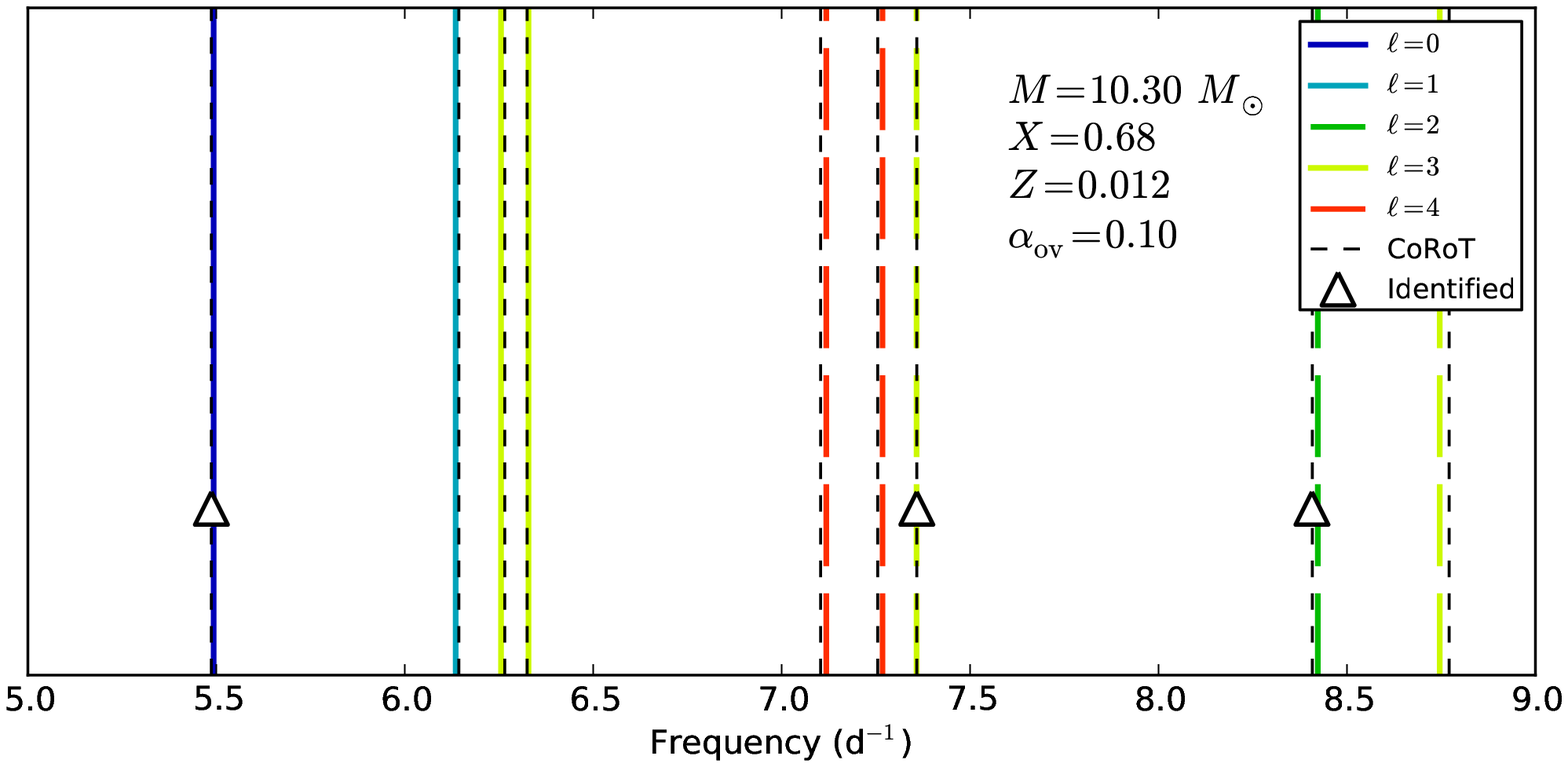}}}
\rotatebox{0}{\resizebox{9cm}{!}{\includegraphics{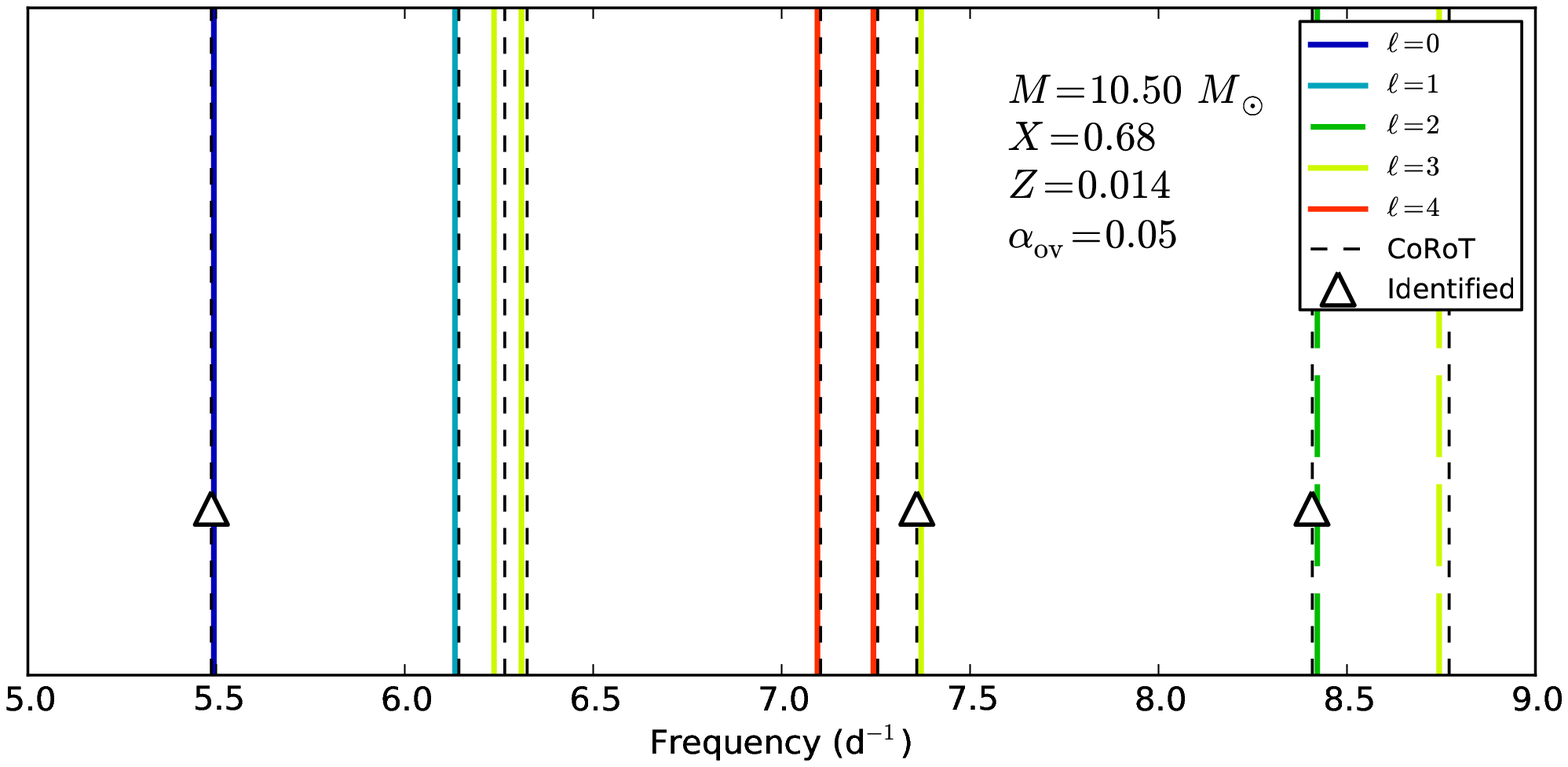}}}
\rotatebox{0}{\resizebox{9cm}{!}{\includegraphics{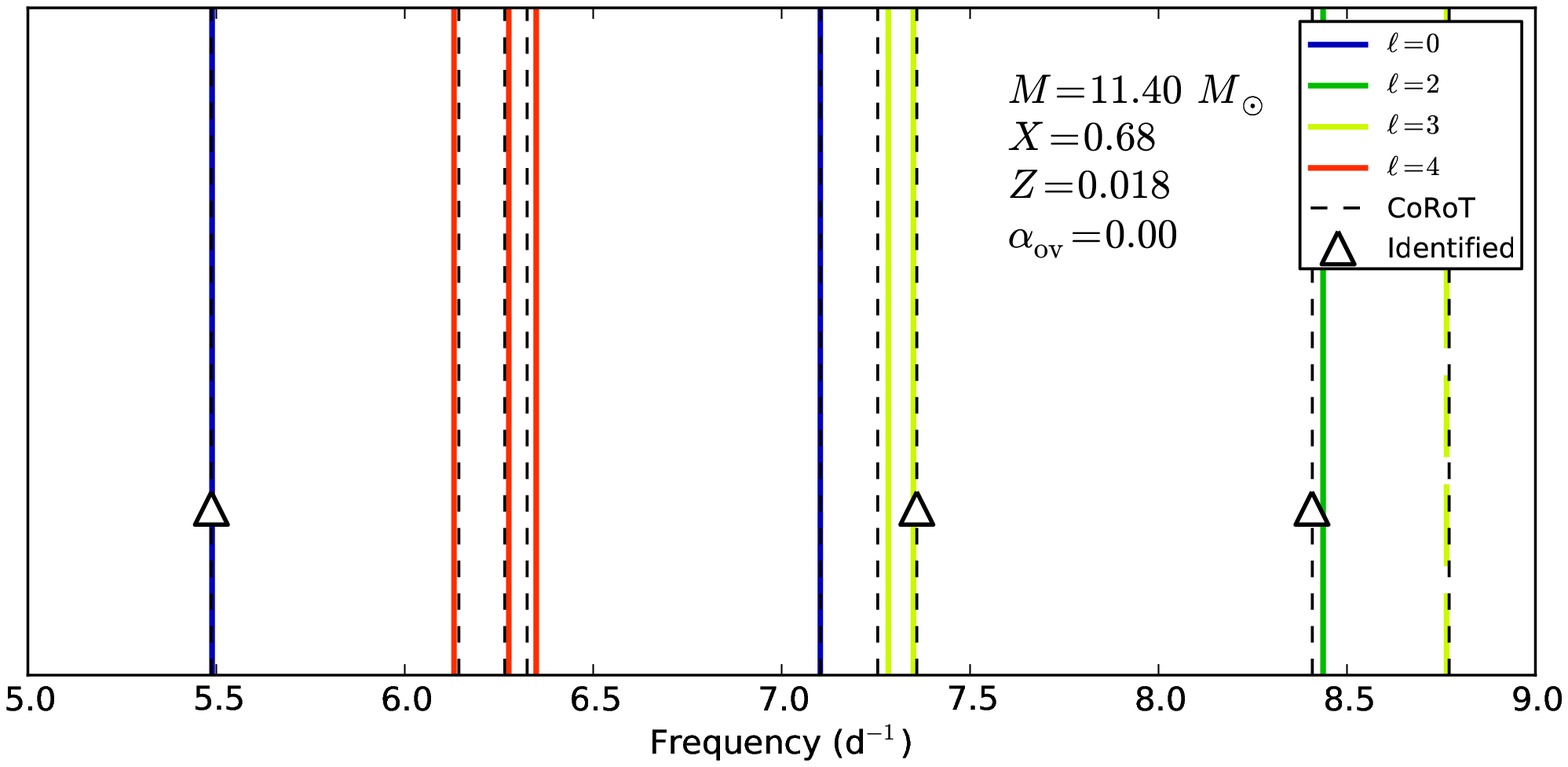}}}
\rotatebox{0}{\resizebox{9cm}{!}{\includegraphics{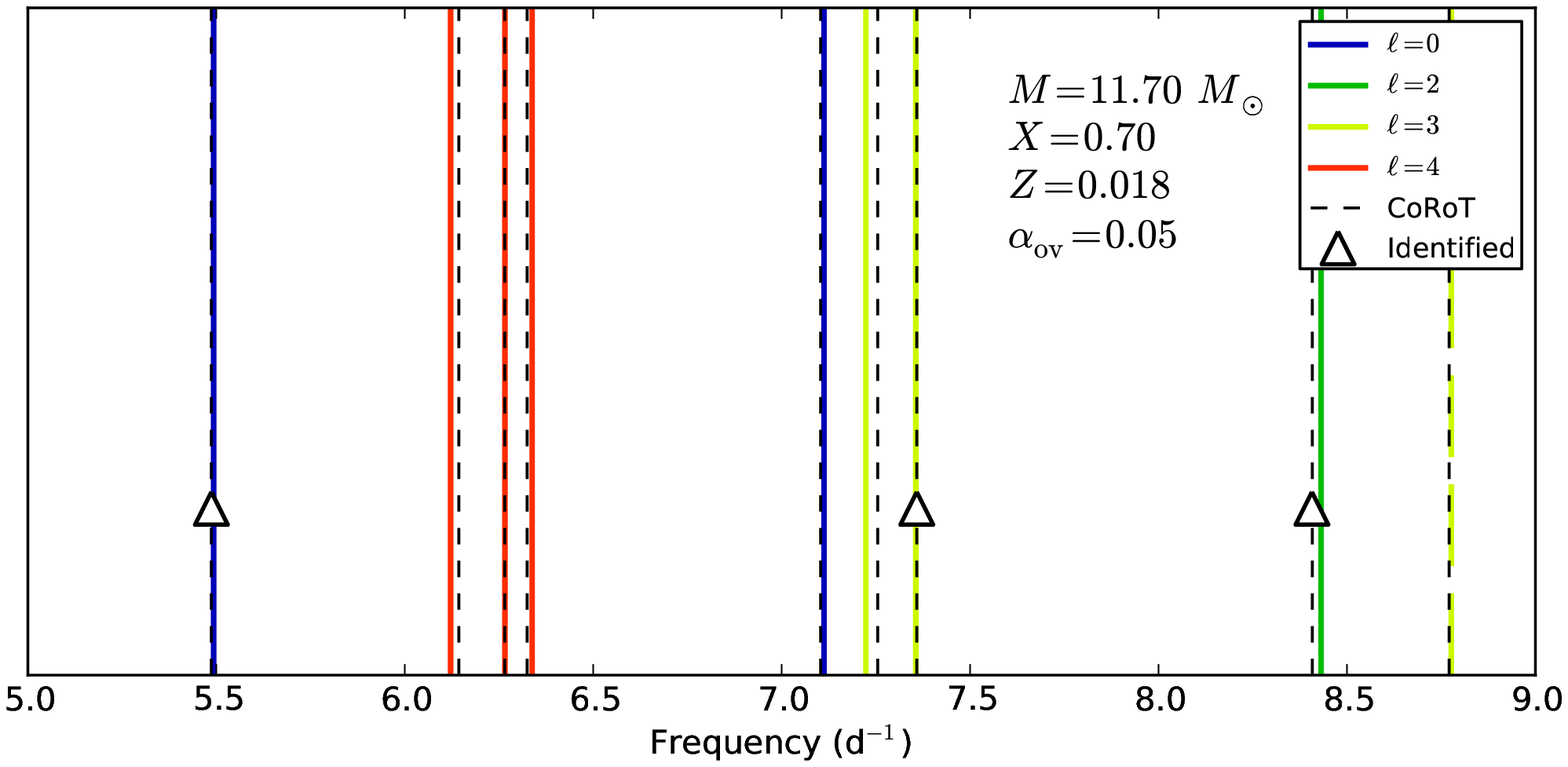}}}
\rotatebox{0}{\resizebox{9cm}{!}{\includegraphics{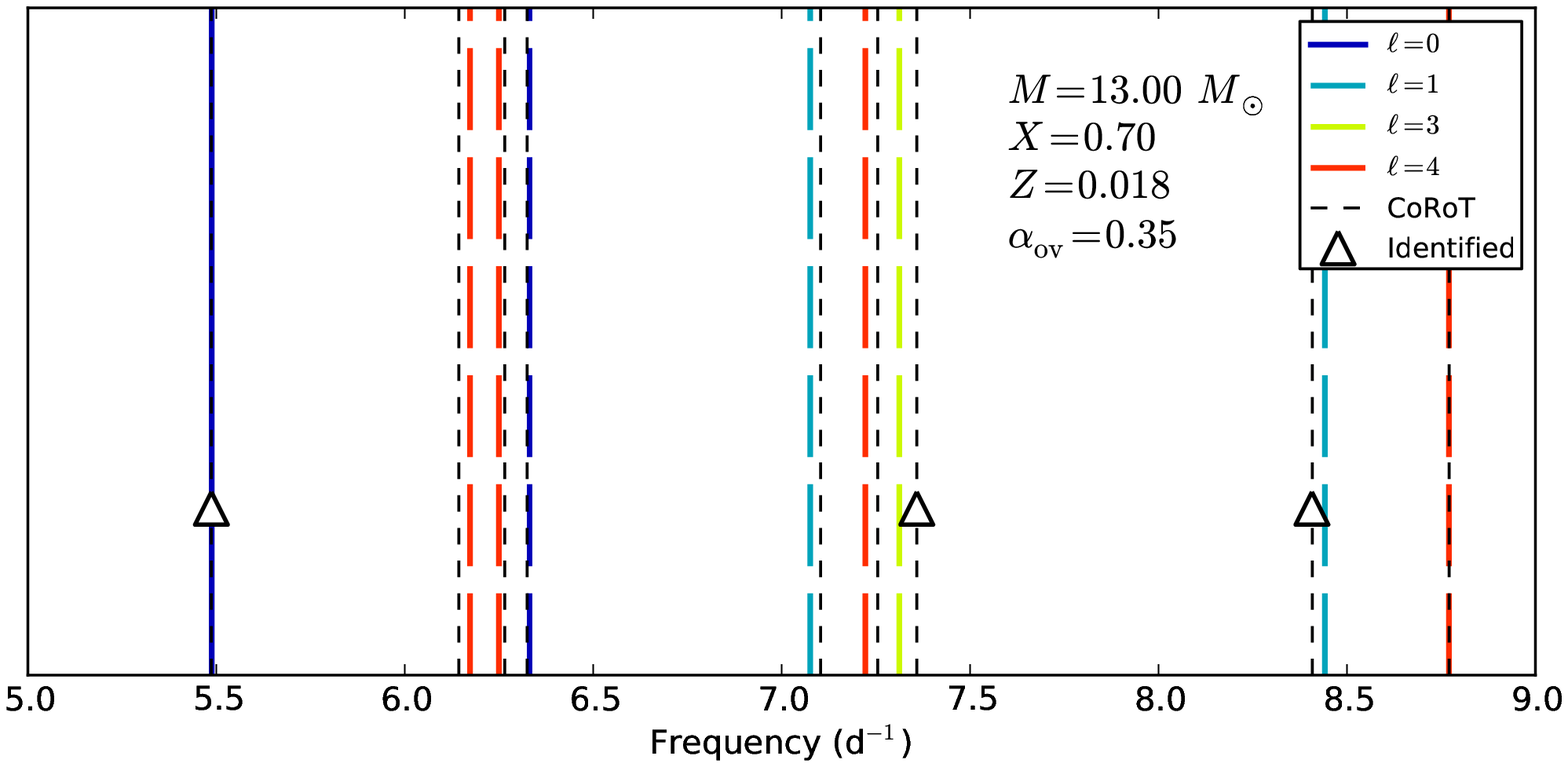}}}
\rotatebox{0}{\resizebox{9cm}{!}{\includegraphics{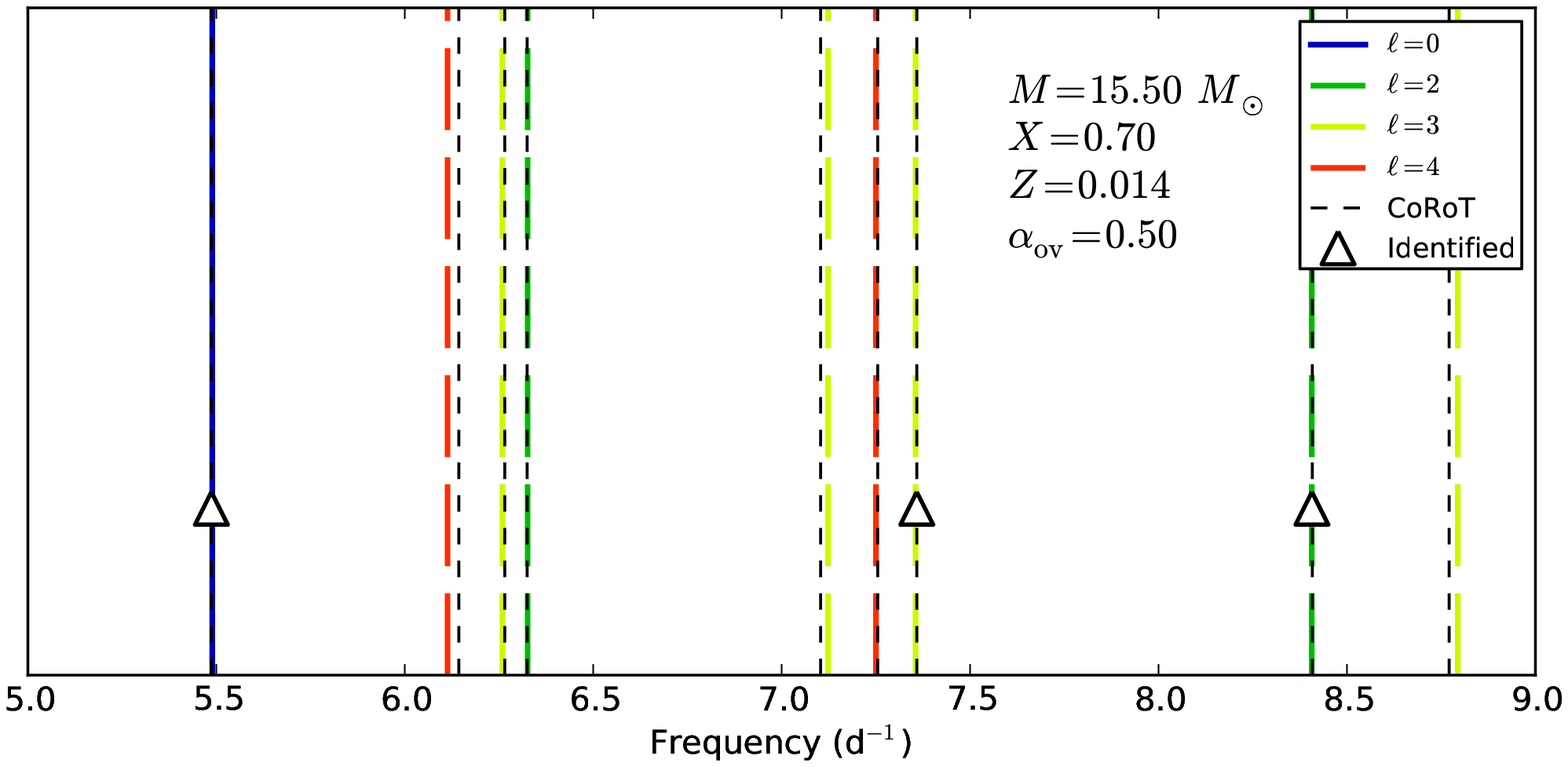}}}
\end{center}
\caption[]{Models with predicted oscillation frequencies 
fitting the nine detected
  frequencies and fulfilling the mode identification as listed in
  Table\,\protect\ref{freqs}. In the four upper panels, the dominant mode is the
  radial fundamental while in the lower left panel it is the first overtone and
  in the lower right panel the fourth overtone. Full coloured lines indicate
  excited modes while dashed coloured lines indicate stable modes.}
\label{bars}
\end{figure*}

\section{Modelling procedure}

The frequencies above 1\,\cd listed in Table\,\ref{freqs} are typical of the 
  low-order p and g modes in $\beta\,$Cep stars (e.g., Aerts et al.\ 2010).  The
  rotation period deduced from the magnetic field measurements combined with
  realistic values for the radius of a $\beta\,$Cep star lead to equatorial
  surface rotation velocities that are low relative to the critical velocity.
Ignoring the effects of rotation in 
  computing evolutionary models and their oscillation frequencies for the
  low-order zonal modes is justified in this situation. 
In practise, this means that we ignore the terms
  proportional to the square of the rotation frequency in all the equations,
including that of hydrostatic support.

For the stellar modelling, we followed a procedure similar to the one adopted by
Ausseloos et al.\ (2004) but for an updated and more extensive grid of 
models. The latter was computed with the Code Li\'egeois d'\'Evolution
Stellaire ({\tt CL\'ES}, Scuflaire et al.\ 2008a), using the input physics
described in Briquet et al.\ (2011) which is not repeated here.  An
extensive evolutionary model grid covering the entire core-hydrogen burning
phase along 
the main sequence (MS), i.e., from the zero-age main-sequence (ZAMS) to
the terminal-age main-sequence (TAMS), for masses $M$ from 7.6\,M$_\odot$ to
20.0\,M$_\odot$ in steps of 0.1\,M$_\odot$, hydrogen mass fractions $X$ from
0.68 to 0.74 in steps of 0.02, metallicities $Z$ from 0.010 until 0.018 in steps
of 0.002 and core overshooting parameters $\alpha_{\rm ov}$ between zero and 0.5
in steps of 0.05 (expressed in local pressure scale heights) were considered. The
probability of catching the star beyond the TAMS, i.e., in the Hertzsprung gap,
is so low that it is not necessary to consider hydrogen-shell burning models.
For each of the MS models along the 27\,500 tracks, a theoretical oscillation
frequency spectrum ignoring the effects of rotation, $\nu_{n\ell}$, was computed
in the adiabatic approximation with the code {\tt LOSC} (Scuflaire et al.\
2008b), covering the range of the low-order (typically up to $n=7$) p and g
modes of degree from $\ell=0$ up to $\ell=4$.

Following Ausseloos et al.\ (2004), we did not restrict the model parameters
{\it a priori\/} by imposing limits derived from classical spectroscopy, because
discrepancies between the seismically and spectroscopically derived fundamental
parameters have been reported for a few massive stars in the literature, in
particular for the gravity (e.g., Briquet et al.\ 2007 for the
$\beta\,$Cep star $\theta\,$Oph and Briquet et al.\ 2011 for the O9V-type
pulsator HD\,46202). Hence, we instead used 
the spectroscopic information as {\it a
  posteriori\/} constraints, after having fitted the nine highest frequencies
listed in Table\,\ref{freqs}.

In the following, we considered models able to fit the dominant radial mode
frequency, after which we continued with those for which this mode is excited.
For the surviving models, we then required that the additional frequencies
higher than the dominant mode frequency be fitted as well.  Next, we considered
the models fulfilling all these requirements, as well as the constraint on the
observed effective temperature. Finally, we also took into account the
excitation of the modes in the frequency range $[5,9]\,$\cd.

\subsection{Fitting the dominant radial mode frequency}

In an initial step, we screened each evolutionary track $(M,X,Z,\alpha_{\rm
  ov})$ from the ZAMS to the TAMS to find two consecutive models whose radial
modes of a particular order encompass the observed frequency of 5.48694\,\cd.  We
considered all pairs of consecutive models that explain the dominant mode
as the fundamental or up to the fourth overtone.

For each of those five paired models along the tracks that fulfil this basic
requirement, we subsequently computed additional models with a smaller time step
to produce one model fitting exactly the frequency of the dominant
radial mode, thereby fixing the age, or equivalently, the effective temperature,
for that combination of $(M,X,Z,\alpha_{\rm ov},n)$. In this way, we ended up
with some 55000 models matching the dominant frequency. These are roughly
equally distributed over all the considered radial orders $n$.

For the seismically modelled $\beta\,$Cep stars with a radial mode reported
in the literature, this mode was either
assumed to be the fundamental or first
overtone, or higher overtones were ruled out because they were not compatible
with the effective temperature or the gravity of the star.
Here, we are dealing with a star whose spectroscopic gravity was
determined to be lower than the one of all those previously modelled stars
($\log\,g=3.45$). Hence, we did not place any limits on the overtone when
interpreting of the dominant mode at this stage because we might be dealing
with a higher overtone if the star is more evolved than usual for $\beta\,$Cep
stars.

\subsection{Requesting mode excitation for the dominant mode}

The oscillations of the $\beta\,$Cep stars are in general understood in
terms of a heat mechanism active in the partial ionisation zones of iron-group
elements 
(Moskalik \& Dziembowksi 1992).
While the majority of detected and well-identified low-order p and g
modes is predicted to be excited by non-adiabatic oscillation computations, some
excitation problems occurred and remain to be solved for a few well-established
observed non-radial mode frequencies, notably the $\ell=1$, p$_2$ modes of the
stars $\nu\,$Eri (Pamyatnykh et al.\ 2004) and $\gamma\,$Peg (Handler et al.\
2009, Zdravkov \& Pamyatnykh 2009). To excite these modes in the
models, one would need to increase the opacities of iron-group elements. In
addition, the predicted frequency of this $\ell=1$, p$_2$ mode for appropriate
models was found to be shifted with respect to the observed value for
both $\nu\,$Eri and $\gamma\,$Peg.  A similar excitation problem occurred for
the star 12\,Lac (Dziembowski \& Pamyatnykh 2008; Desmet et al.\ 2009).

Theory explains the excitation of the observed radial and p$_1$ non-radial 
modes well for the case studies of seismically modelled
$\beta\,$Cep stars, except for the very massive ($M\simeq
24\,$M$_\odot$) O9V pulsator HD\,46202, for which none of the detected
frequencies is predicted to be excited (Briquet et al.\ 2011). 
In particular, the detected radial modes
of {\it all\/} seismically modelled $\beta\,$Cep stars of spectral type B 
agree with excitation predictions for these particular modes. We are therefore
justified in insisting that the dominant mode of HD\,180642 be excited in
present-day theoretical computations.  To use this requirement as an additional
constraint in our stellar modelling, we performed non-adiabatic computations for
the radial modes of all the $\sim$55\,000 selected models discussed in the
previous section with the code {\tt MAD\/} (Dupret 2001, Dupret et al.\ 2002).

For $\sim$18\,850 of the models that fit the frequency 5.48694\,\cd as a radial
mode, this mode is predicted to be excited. The distribution over the various
overtones is as follows: 9964 models have the dominant mode excited as
fundamental, 6681 as first overtone, 892 as second overtone, and 1324 as fourth
overtone.  There are no models with the third overtone excited while there are
with an excited fourth overtone.  This may seem strange at first sight, but is
explained by the occurrence of a resonance of the type $\omega_{{\rm
    p}_4}\simeq\,2\omega_{{\rm p}_1}$. It was already known that a resonance
$\omega_{{\rm p}_6}\simeq\,3\omega_{{\rm p}_1}$ occurs in polytropic models (Van
Hoolst 1996). Near-resonances were also found in that study, in particular the
one we also find here on the basis of more realistic models (Fig.\,9 in Van
Hoolst 1996).
In all the models with the fourth overtone excited, covering the mass
interval $[16.3,20]\,$M$_\odot$, we also found the fundamental mode at 2.74\,\cd
(and quite often also the first overtone) to be excited. This frequency does not
occur among the 127 dominant frequencies of HD\,180642. 
These high-overtone models are more evolved and in
closer agreement with the spectroscopic $\log\,g$ than the models with the
fundamental mode interpretation.

\subsection{Matching the additional non-radial mode frequencies}

In the subsequent steps of the modelling procedure, we could not rely on a unique
$(\ell,m)$ identification of the modes corresponding to the observed
frequencies, as only limited additional information is available (see
Table\,\ref{freqs}).  Moreover, the additional modes all have much lower
amplitudes than the one of the dominant mode, thus we did not prefer one
above the other for the matching procedure. Thus we requested at once that all
eight additional frequencies listed in Table\,\ref{freqs} were fitted by model
frequencies.

We also imposed that the identification of the degree of the mode with frequency
7.35867\,\cd is fulfilled ($\ell$=0 or 3). It turned out that this frequency is
not a higher overtone radial mode.  For the frequency 8.40790\,\cd, we requested
it to have one of the seven $(\ell,m)$ combinations listed in
Table\,\ref{freqs}. For each of those possibilities for $(\ell,m)$, we used the
rotational splitting 
along with the $V_{\rm eq}$ value listed in  Table\,\ref{freqs}
and model
radius to deduce the shift in mode frequency implied by rotation, according to
the formula
\begin{equation}
\displaystyle{
\nu_{n\ell m} = \nu_{n\ell 0} + \left(m\ \beta_{n\ell}\ \frac{V_{\rm eq}}{2\pi R}
  \times 86\,400\right)\,{\rm d}^{-1}}
\label{ledoux}
\end{equation}
(e.g., Aerts et al.\ 2010, Chapter 3), i.e., we limited the frequency
  matching procedure to first-order effects in the rotation frequency, which is
  consistent with our grid of spherically-symmetric evolutionary models.

  In the matching procedure, we screened all the models which excite and fit the
  dominant radial mode as described in the previous section and fulfilling all
  the information in Table\,\ref{freqs} for the eight additional measured
  frequencies, after shifting them according to Eq.(\ref{ledoux}) and
  Table\,\ref{freqs}.  The quality of the frequency matching was evaluated by
  the computation of a reduced $\chi^2$ statistic, where we used 0.02\,\cd as
  the error for all the eight frequencies because it is a typical uncertainty on
  theoretically predicted frequencies for non-rotating main sequence pulsators
  due to differences in the input physics of the models (e.g., Moya et al.\
  2008). We then considered only those models with $\chi^2<1$ and for which each
  individual frequency is fitted to better than 0.04\,\cd. The latter
  constraint was adopted to avoid a rather large mismatch 
  leading to $\chi^2<1$.

  In total, 2541 models survived our stringent matching constraints, covering a
  mass range of $[8.1,19.9]\,$M$_\odot$.  Among those, some 1000 predict that
  the dominant frequency is the fundamental mode, some 1200 that it is instead
  the first overtone and a few hundred that it is either p$_3$ or p$_5$.

  Six prototypical examples of schematic frequency spectra are shown in
  Fig.\,\ref{bars}. The four that fit the dominant mode as the radial
  fundamental, predict that the 8.4\,\cd frequency is an $\ell=2, m=+1$ mode and
  that the star has an equatorial rotation velocity of 26\,km\,s$^{-1}$. In the
  case of the first overtone model for the dominant mode, the 8.4\,\cd is an
  $\ell=+m=1$ mode for an equatorial rotation of 119\,km\,s$^{-1}$, and in case
  of the fourth overtone model, the 8.4\,\cd is an $\ell=+m=2$ mode for a
  rotation velocity of 52\,km\,s$^{-1}$. As can be seen from Fig.\,\ref{bars},
  all those cases give a good explanation of the observed frequencies. We stress
  that these are only six out of many good solutions.

\begin{figure}
\begin{center}
\rotatebox{270}{\resizebox{6.5cm}{!}{\includegraphics{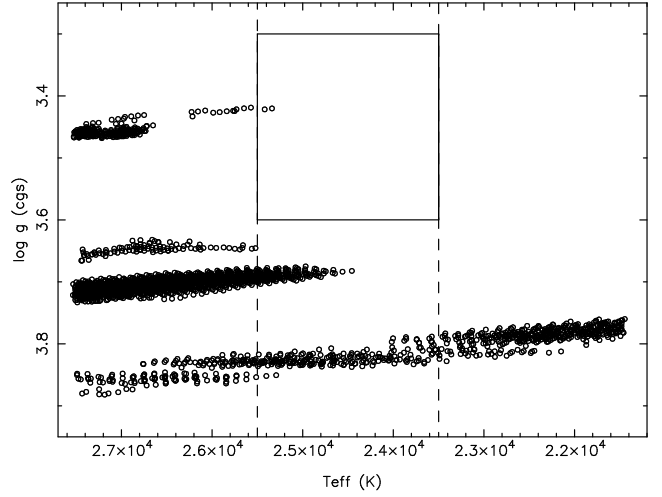}}}
\end{center}
\caption[]{Kiel diagram showing all the models with a radial mode at the
  frequency 5.48694\,\cd excited and matching the additional eight higher mode
  frequencies listed in Table\,\protect\ref{freqs}. The $1\sigma$ error box for
  the spectroscopically derived $T_{\rm eff}$ and $\log\,g$ is shown as well.
  The dashed vertical lines indicate the restriction we imposed on the models.}
\label{kiel}
\end{figure} 

\begin{figure}
\begin{center}
\rotatebox{270}{\resizebox{6.5cm}{!}{\includegraphics{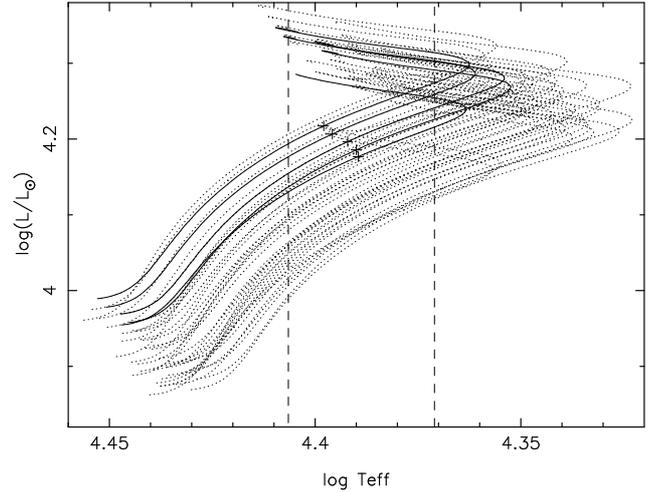}}}
\end{center}
\caption[]{HR\,diagram showing the five models listed in
the upper part of
  Table\,\protect\ref{models} as crosses 
on their evolutionary track (full lines).
  These models
  have excited oscillation modes whose frequencies match all the observed ones,
  except for the frequency 8.77\,\cd, which fits but is predicted to be stable.
  The 38 dotted tracks each contain one model with excited oscillation
  frequencies matching the observed ones, except for the two modes with
  frequency above 8\,\cd, which are predicted to be stable.}
\label{hrd}
\end{figure} 

We point out that our procedure was not ideal for the three options for
8.40790\,\cd with the highest equatorial rotation velocity listed in
Table\,\ref{freqs}, for which second-order rotational effects may come into play
at a level that can reach values above the theoretical uncertainty we adopted.
The estimate of the rotation period from the magnetic field study by Hubrig et
al.\ (2011) would have allowed us to drop these three options from the start.
We preferred not to do so, however, as the derivation of the rotation period by
Hubrig et al.\ relies on the assumption of dealing with a rigid rotator model
and a centred dipole. While this is a very plausible assumption, and certainly
the most obvious one to make, we wanted the seismic modelling to be as
independent as possible of this result. This implies that our modelling is not
optimal for those three options with high $V_{\rm eq}$, because we did not
consider second-order effects in the rotation when computing the evolutionary
models and the pulsation frequencies.  Improving this would require a full
two-dimensional treatment of the equilibrium models and their frequencies, which
is beyond the scope of this paper.

\subsection{Constraints imposed by spectroscopy}

There has been good agreement between spectroscopically derived effective
temperatures and abundances, and their seismically derived counterparts, from
the method of frequency matching that we adopted here. This is not necessarily
the case for the gravity of the seismically modelled $\beta\,$Cep stars. While
good agreement was found for several stars (e.g., HD\,129929, Dupret et al.\
2004; 12\,Lac, Desmet et al.\ 2009), the spectroscopic $\log\,g$ turned out to
be 0.15 dex higher than the seismic $\log\,g$ for the star $\theta\,$Oph
(Briquet et al.\ 2007). For HD\,46202, the discrepancy even amounted to 0.24 dex
(Briquet et al.\ 2011).  Given these previous discrepancies, we prefer not to
impose the spectroscopic 
gravity as a secure constraint in the modelling; this would imply that
the dominant mode is a high overtone ($n\geq 3$) while we consider this
unlikely. Similarly, the current metallicity of the star measured at its surface
is not necessarily the initial one when the star was born, as effects such as
atomic diffusion or unknown mixing may have occurred. Thus, we take a
conservative attitude and request as an additional constraint from spectroscopy
that the models must fulfil the $1\sigma$ error bar in the effective temperature
only.

In Fig.\,\ref{kiel}, all 2541 models surviving the previous section are shown,
along with the spectroscopically determined error box in the Kiel diagram. As
can be seen, only two models would survive, if we were to impose the $1\sigma$
error bar for the gravity of the star. For these two models, none of the observed
non-radial modes is excited (see the following section). We consider this too
restrictive and continued with all the models that fulfil $T_{\rm eff} \in
[23\,500,25\,500]$K. There are 357 models left after this requirement, with a
mass coverage from 9.5\,M$_\odot$ to 15.6\,M$_\odot$. This set of models
still covers the entire considered range of values for the metallicity, hydrogen
fraction, and core overshooting.

\subsection{Excitation of the non-radial modes}
\begin{table*}
\caption{Stellar models in agreement with 
the observational properties of HD\,180642.}
\label{models}
\centering
\begin{tabular}{cccccccccc}
\hline
Mass  & $Z$ & $\alpha_{\rm ov}$ & $X$ & $T_{\rm eff}$ & 
$\log\,g$ & $\log (L/L_\odot)$ & R & Age & $X_c$ \\
 (M$_\odot$) & & & & (K) & (cgs) & & (R$_\odot$) &($10^8$\,yr) & \\
\hline
{\it 11.4} &0.018 &0.00 &0.68 &24520 &3.829 &4.177&  6.80& 0.1243& 0.209\\
11.6 &0.018 &0.05 &0.68 &24880 &3.832 &4.206&  6.84& 0.1220& 0.233\\
11.7 &0.018 &0.05 &0.68 &25000 &3.834 &4.217&  6.86& 0.1200& 0.235\\
{\it 11.7} &0.018 &0.05 &0.70 &24550 &3.834 &4.186&  6.86& 0.1296& 0.246\\
11.8 &0.018 &0.05 &0.70 &24670 &3.835 &4.197&  6.88& 0.1275& 0.248\\
\hline
10.5 & 0.014& 0.05 & 0.68 & 23880& 3.814 & 4.111 & 6.64 & 0.1482 & 0.182\\
\vdots & \vdots & \vdots & \vdots & \vdots & \vdots & \vdots & \vdots & \vdots & 
\vdots \\
12.0 & 0.018 & 0.05 & 0.72 &  24460 & 3.837 & 4.187 & 6.91 & 0.1332 & 0.262\\ 
\hline
\end{tabular}
\end{table*}

We subsequently checked the mode excitation of the low-order zonal frequencies
for the 357 surviving models with the code {\tt MAD}.  Depending on the model
parameters, we found excitation for the first radial overtone near frequency
7.1\,\cd for masses between 9.7 and 13\,M$_\odot$, for the $\ell=1$, p$_1$ mode
with frequency in $[6.0,6.5]\,$\cd for masses between 9 and 11\,M$_\odot$, and
for the $\ell=2$ f mode with frequency near 6.9\,\cd for masses between 9.4 and
11.6\,M$_\odot$. For $\ell=3$ and $\ell=4$, the g$_1$ modes also have their
frequency in $[6.0,6.5]\,$\cd and are found to be excited for masses between 9
and 12\,M$_\odot$ while the $\ell=3$ f mode has its frequency near 7.35\,\cd and
is excited for masses between 9.7 and 12.2\,M$_\odot$. These rough conclusions
about the mode excitation are in qualitative agreement with the results on mode
excitation found by Dziembowski \& Pamyatnykh (2008), but these authors
considered only a few representative models that have frequency spectra similar
to those observed for the $\beta\,$Cep stars $\nu\,$Eri and 12\,Lac, while we
considered an entire grid of models.

To constrain the parameters space for HD\,180642, we first eliminated the models
for which none of the non-radial modes with frequency above that of the dominant
radial mode is excited.  This reduced the number of models to 221 and led to an
upper limit in mass of 14.1\,M$_\odot$. In particular, this removed the two
models with the fourth overtone as a fit to the dominant frequency. Only five
models able to predict that the dominant mode is the first overtone
remained. All of these have only one non-radial mode excited in the frequency
range $[6,9]\,$\cd. In each case, this corresponds to the $\ell=2$, p$_1$ mode
near 6.2\,\cd. This is insufficient to explain the observed frequency spectrum
and we are thus left with models whose fundamental radial mode fits the dominant
frequency.

Among the remaining 221 models, we found five that excite {\it all but the
  highest frequency\/} listed in Table\,\ref{freqs}. The properties of these
five models are listed in the upper part of Table\,\ref{models}.  These models
are indicated with a cross in Fig.\,\protect\ref{hrd} and fulfil the
spectroscopically determined effective temperature. The frequency spectra of the
two models whose mass is indicated in italics in Table\,\ref{models} were
compared with the observed ones in the middle panels of
Fig.\,\protect\ref{bars}.  Given that excitation problems occur for some modes
with frequencies above 7\,\cd in the three well-known bright class members
$\nu\,$Eri, 12\,Lac, and $\gamma\,$Peg, it is quite encouraging to find models
whose oscillation spectra are in full agreement with all the numerous
observational constraints, except for the spectroscopic gravity and the
excitation of the $\ell=3$, p$_1$ mode.

In addition to the five remaining models, there are 38 models for which the mode
excitation is fulfilled, except for the two highest frequencies in
Table\,\ref{freqs}, i.e., for one frequency less than the five models in the
upper part of Table\,\ref{models}. The evolutionary tracks to which these also
``acceptable'' models in the lower part of Table\,\ref{models} belong are
indicated as thin dotted lines on Fig.\,\ref{hrd}, for reference with respect to
the five best models. These 43 models together cover a mass range of
$[10.5,12.0]$\,M$_\odot$, the full range $X\in [0.68, 0.74]$ of the grid, $Z\in
[0.014, 0.018]$, and $\alpha_{\rm ov}$ between zero and 0.2. That the initial
internal metallicity of the star seems to be somewhat higher (1$\sigma$ level)
than the observed one at the surface for all the acceptable models, might have
been introduced by the discrepancy in the gravity, which is not treated
independently from the microtubulence in a spectroscopic analysis.  It might
also be due to effects of diffusion given HD\,180642's longitudinal
magnetic field of up to 700\,G (Hubrig et al.\ 2009, 2011).

Almost all the seismic models have $\log\,g\simeq3.83$, which is
0.38\,dex (2.5$\sigma$) above the best estimate from a classical spectroscopic
analysis (Paper\,II).  We thus face another case where the seismic gravity
differs from the spectroscopic gravity, just as for the
$\beta\,$Cep pulsators $\theta\,$Oph (B2IV, Briquet et al.\ 2007) and 
HD\,46202 (O9V, Briquet et al.\ 2011).  Combining the seismic gravity with the
equatorial surface rotation velocity of 26\,km\,s$^{-1}$ and a radius of
$R\simeq\,6.8\,$R$_\odot$, leads to a surface rotation frequency of 0.075\,\cd
(rotation period of 13.3\,d). We thus find that a factor of 
four occurs between the
surface rotation frequency and the lowest frequency that was jointly detected
in the CoRoT light curve and the spectroscopic time series. 
The frequency 0.89870\cd listed in Table\,\ref{freqs} is 12 times the
  rotation frequency, within the measurement errors.
The seismic rotation period that we obtained from our best-fitting models is in
excellent agreement with the most probable period deduced from the magnetic field
measurements.

\subsection{The full oscillation spectrum}
\begin{figure*}
\begin{center}
\rotatebox{270}{\resizebox{11cm}{!}{\includegraphics{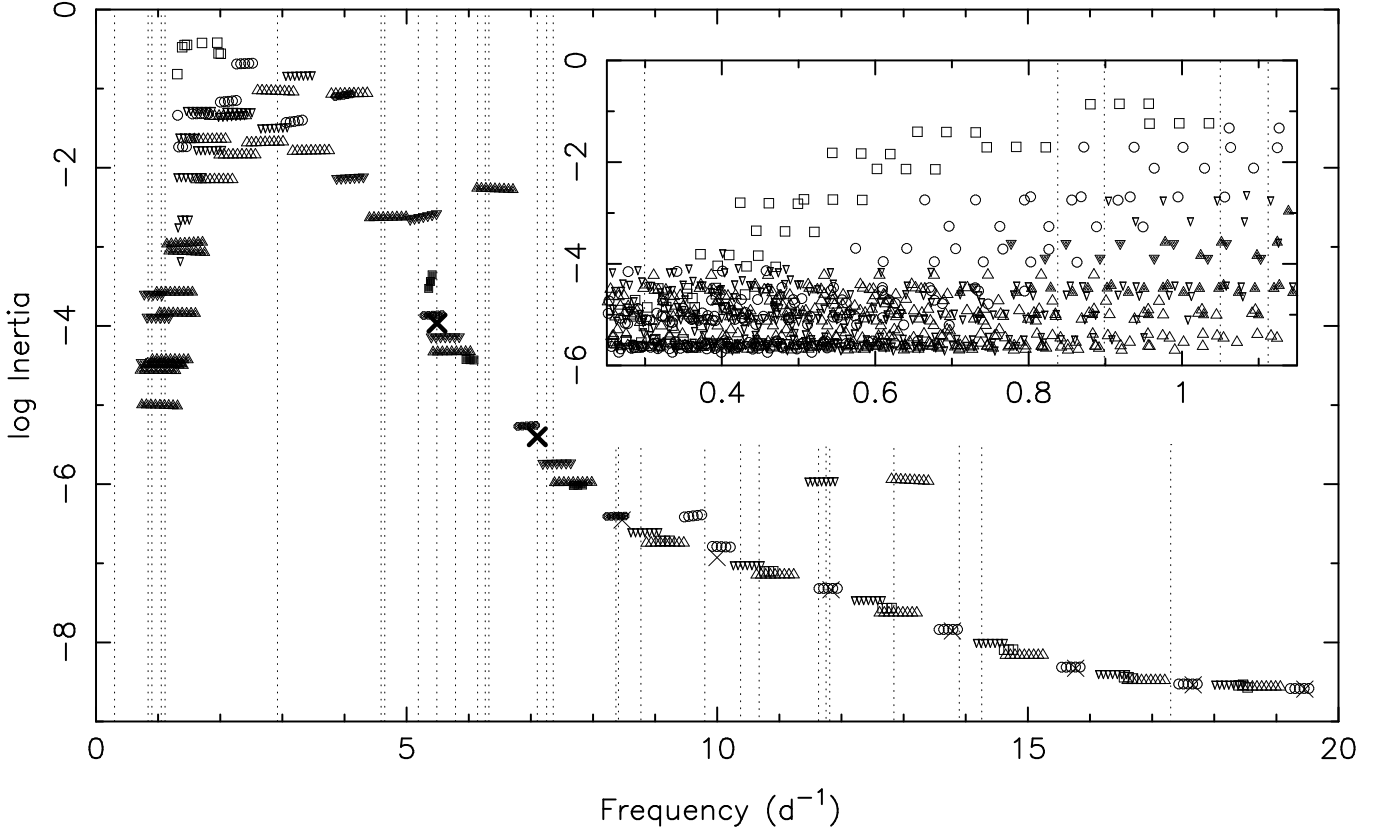}}}
\end{center}
\caption[]{Mode inertia for the oscillations of the model with
  $M=11.4\,$M$_\odot$, whose other parameters are listed in
  Table\,\protect\ref{models}. Thick crosses indicate the radial fundamental and
  first overtone, which are both excited. Thin crosses indicate the higher
  overtone radial modes, which are not excited.  Filled symbols indicate excited
  non-radial modes and empty symbols stable modes according to the indicated
  label.  
The meaning of the symbols is as follows: cross: $\ell=0$, squares:
  $\ell=1$, circles: $\ell=2$, downward triangle: $\ell=3$, upward triangle:
  $\ell=4$. 
For clarity, we show only the excited g modes in the region below
  1.3\,\cd in the large panel, while all g modes are shown in the inset.  The
  dotted vertical lines indicate the frequencies detected in the CoRoT light
  curve (Paper\,I), nine of which have also been detected in ground-based
  data (Paper\,II).}
\label{inertia}
\end{figure*}
 \begin{figure}[h!]
\begin{center}
\rotatebox{0}{\resizebox{9cm}{!}{\includegraphics{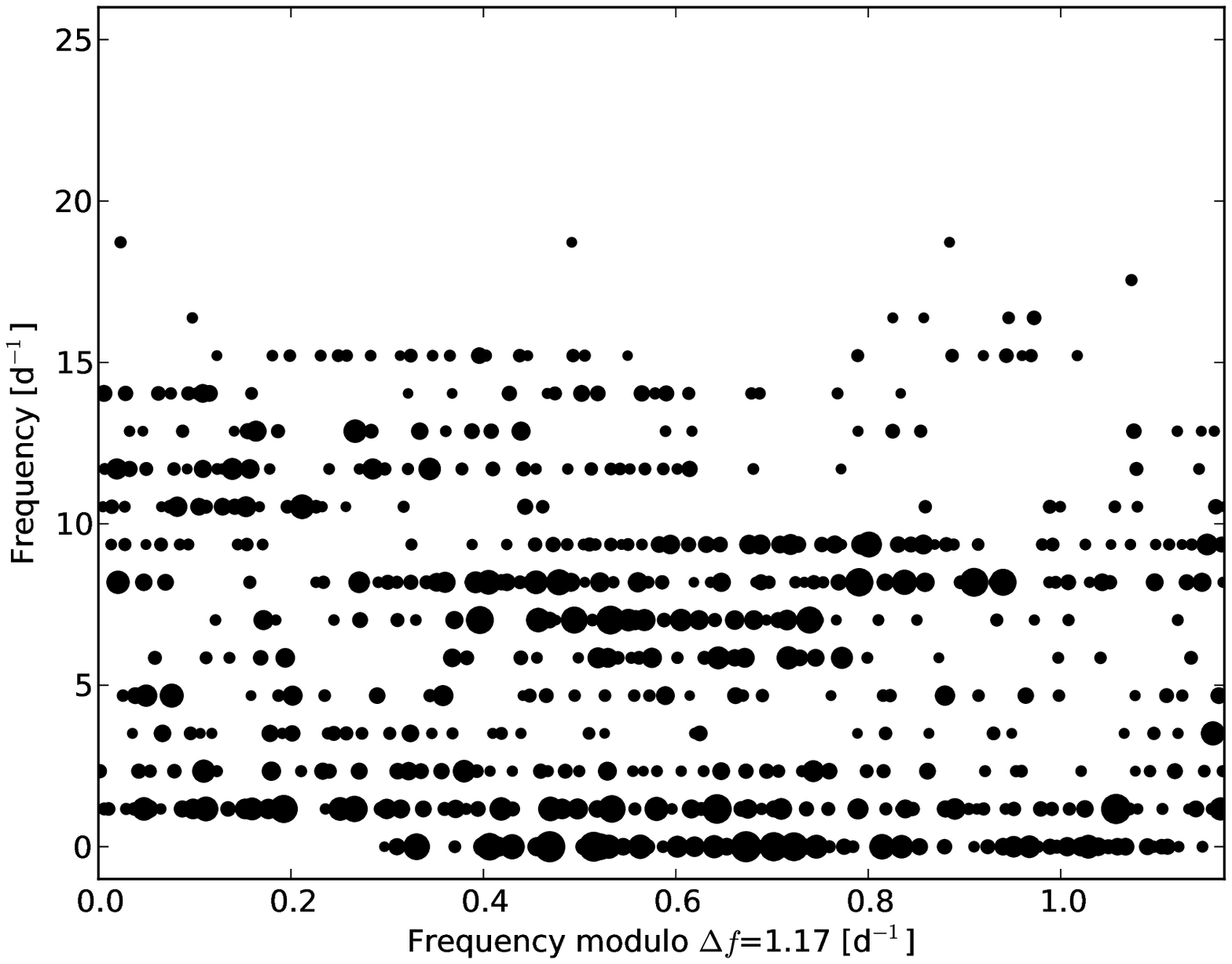}}}
\rotatebox{0}{\resizebox{9cm}{!}{\includegraphics{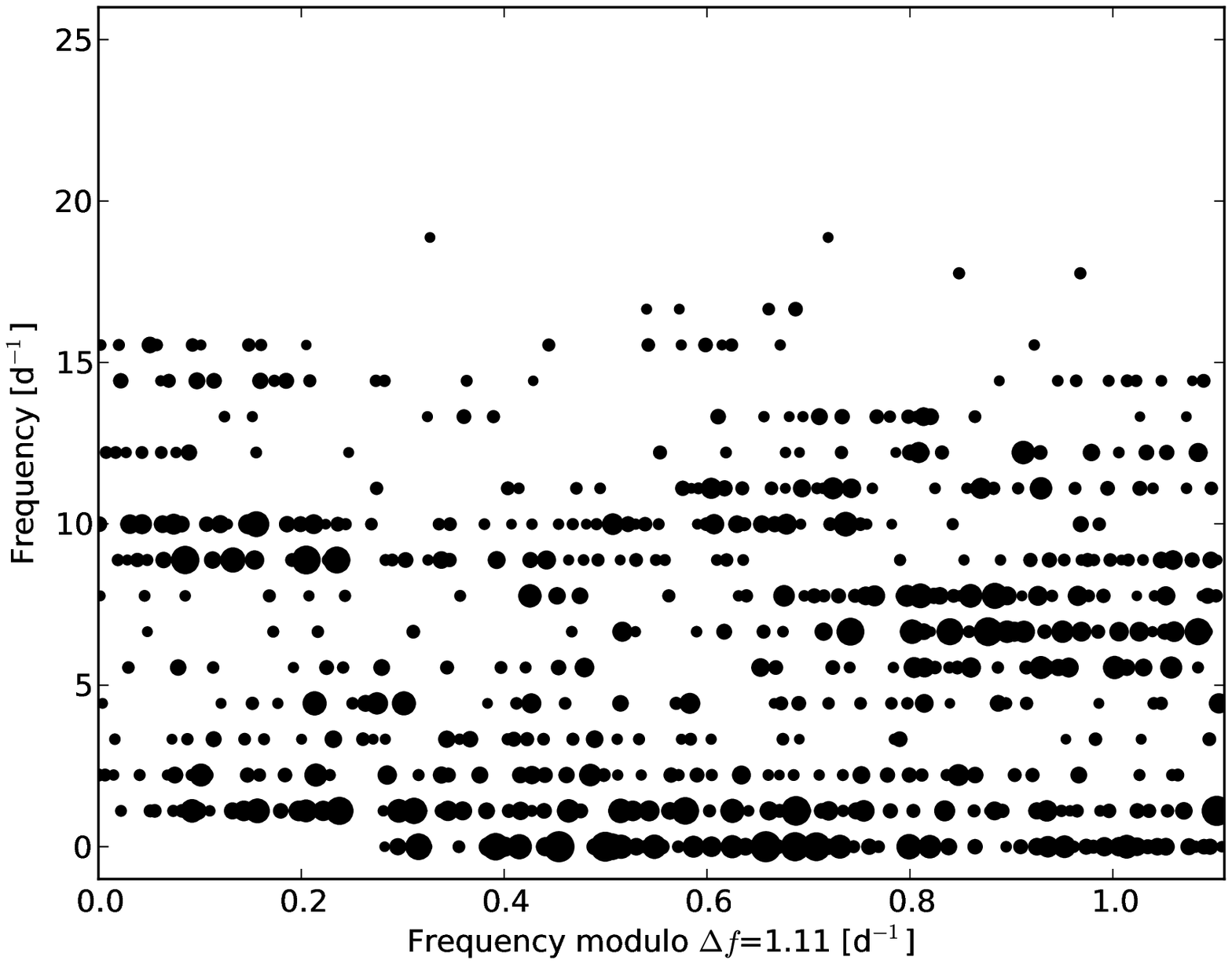}}}
\rotatebox{0}{\resizebox{9cm}{!}{\includegraphics{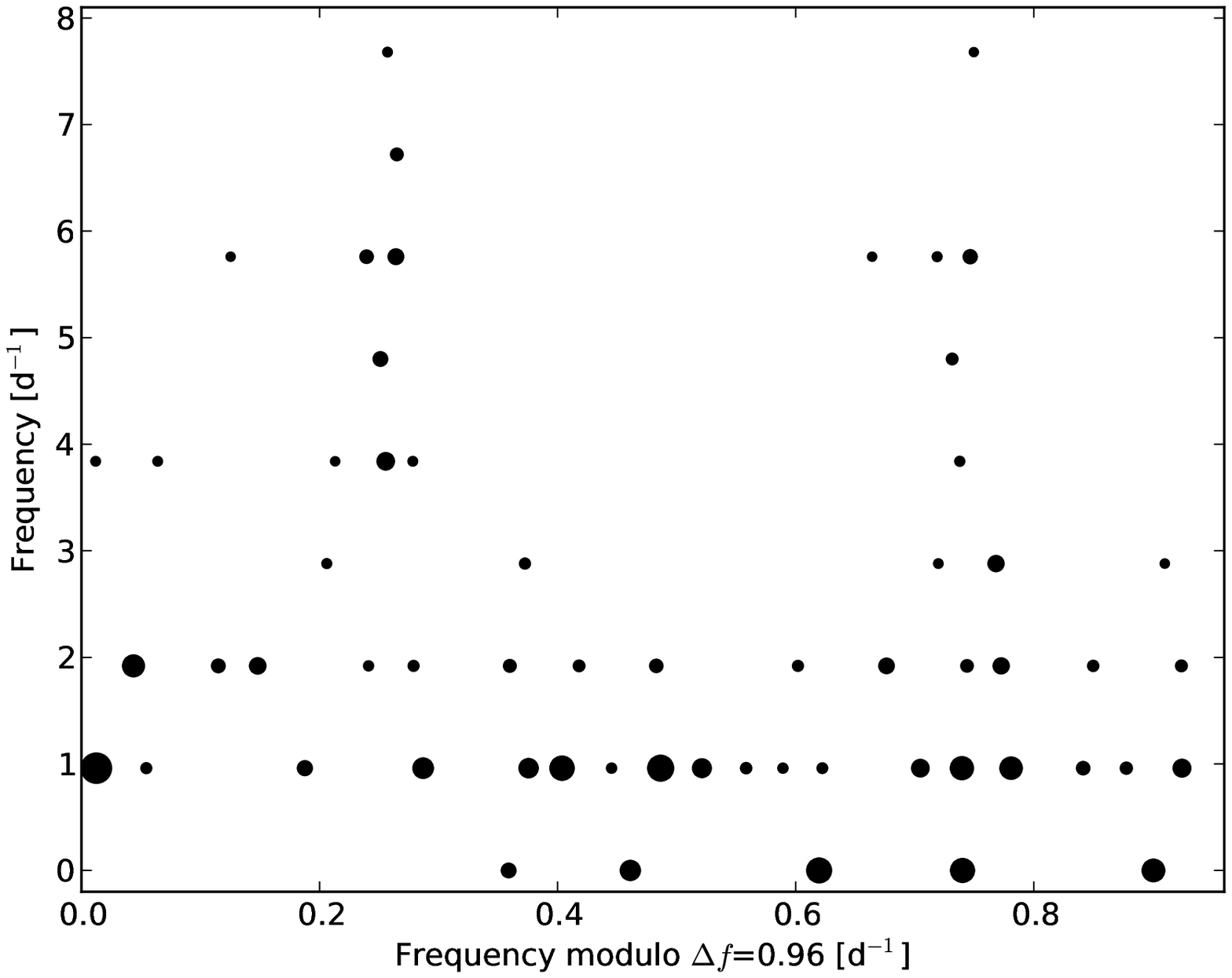}}}
\end{center}
\caption[]{\'Echelle diagrams of the CoRoT frequencies for HD\,180642 for the
  large spacing determined by Belkacem et al.\ (2009, upper panel) and in
  Paper\,I (middle panel). For comparison, we also show the \'echelle diagram of
  the O8.5V star HD\,46149 reproduced from Degroote et al.\ (2010b), 
where clear ridges produced by stochastically excited modes are found. 
The size of the symbols
scales with the power of the frequency.}
\label{echelle}
\end{figure} 

\begin{figure}
\begin{center}
\rotatebox{270}{\resizebox{4.5cm}{!}{\includegraphics{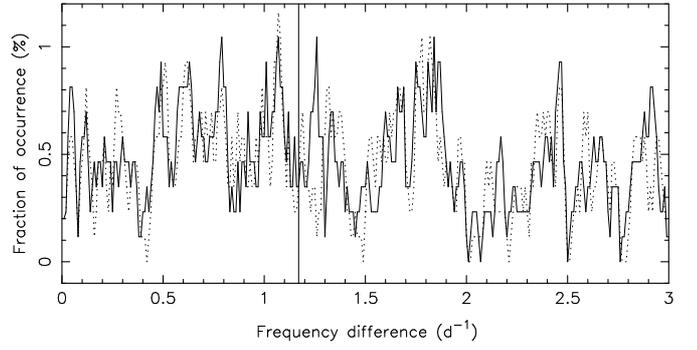}}}
\end{center}
\caption[]{Fraction of the occurrence of values for the frequency differences
  between zonal p modes of degree $\ell=0,\ldots,4$ in the frequency
  regime $[11,26]\,$\cd compared with the spacing of 1.17\,\cd reported by
  Belkacem et al.\ (2009). The full line is for the model of 11.4\,M$_\odot$,
  while the dotted line is for the model with 11.7\,M$_\odot$ indicated in
  italics in Table\,\protect\ref{models}. }
\label{spacingfig}
\end{figure} 

Continuing further with the models in Table\,\ref{models},
we considered
their entire oscillation spectrum in the range $[0.2,40]\,$\cd, 
taking into account all rotational splittings
according to the Ledoux formulation (e.g., Aerts et al.\ 2010) for 
a rotation frequency of 0.075\,\cd, i.e., assuming rigid
rotation. In Fig.\,\ref{inertia}, we show the mode inertia (defined as
Eq.\,(3.140) in
Aerts et al.\ 2010) versus the
oscillation frequencies for the seismic model with $M=11.4\,$M$_\odot$ listed in
Table\,\ref{models} in the range $[0,20]\,$\cd, for modes with $\ell=0,\dots,4$.
We note that the inertia follow a smooth curve as the radial order increases,
except for some specific modes of $\ell=3$ and 4 where the details of the 
trapping in the interior change owing to the interplay between the mode frequency
and the Br\"unt-V\"ais\"al\"a and Lamb frequencies. 
The detected frequencies in the best fitting CoRoT light curve model deduced in
Paper\,I are shown as dotted vertical lines, except for the harmonics of the
dominant frequency. Full symbols denote modes predicted to be excited from the
non-adiabatic computations done with the code {\tt MAD}, while the open symbols
are predicted to be stable.

All the combination frequencies (listed in Table\,2 of Paper\,I) in the range
$[4,8.5]\,$\cd can be closely identified 
with frequencies of excited modes.  This is also
true for the four frequencies in the range $[0.8,1.2]\,$\cd, which can be
explained by excited g modes of intermediate to high radial order ($n$ typically
between 10 and 30). All other combination frequencies found in Paper\,I can be
attributed to 
model frequencies of non-excited p modes taking into account the observational
and theoretical uncertainties. If the frequency 0.3\,\cd were produced by 
pulsation, it
would have to correspond to an $\ell=3$ mode (Paper\,II). There are indeed
various $\ell=3$ modes available, taking into account the rotational splitting
(see the inset of Fig.\,\ref{inertia}). These have radial orders from 53 to 60.
Given the magnetic field detection for the star, it might also be that this
frequency detected in both the photometry and spectroscopy
originates from low-amplitude surface inhomogeneities. We are unable to
discriminate between these two options. In any case, we interpret the detection
of all these modes with combination frequencies that are not excited by the heat
mechanism in terms on a 
non-linear resonant mode excitation by the dominant radial
mode, as already proposed in Paper\,I. That this mechanism seems to be at work
and results in frequencies beyond the usual frequency regime of the $\beta\,$Cep
stars is supported by the low inertia of these modes (Fig.\,\ref{inertia}),
which makes it much easier to excite them than low-order g modes.  The high
amplitude of the radial mode thus seems to trigger the excitation of p modes at
frequencies between $[8.5,20]$\,\cd, which would otherwise be stable.

The results discussed for the model with $M=11.4\,$M$_\odot$ remain
qualitatively the same for the four other models listed in the upper part of 
Table\,\ref{models}.

\subsection{Revisiting the case of solar-like oscillations}
\label{spacing}

We used the set of models that fulfil all the seismic requirements of the
$\beta\,$Cep-type modes, as well as the combination frequencies, to revisit the
claim of solar-like oscillations in the star (Belkacem et al.\ 2009). In
Paper\,I, the authors already pointed out that the value found for a possible
frequency spacing depends on the prewhitening procedure adopted to model the
light curve.  Moreover, it was found in Paper\,I that several of the modes with
combination frequencies considered by Belkacem et al.\ (2009), who made their
computations for the frequency range $[11,26]\,$\cd, are phase-locked, which is
not what is expected for stochastic forcing.  The interpretation of the residual
light curve in terms of solar-like oscillations was also questioned by Balona et
al.\ (2011), in view of the lack of such oscillations in {\it Kepler\/} data of
B-type pulsators, because these data have higher precision than the CoRoT
data. One expects solar-like oscillations in stars with similar fundamental
parameters than HD\,180642 to have comparable amplitudes, and these amplitudes
would be far above the detection limit of {\it Kepler\/} data. In the sample of
Balona et al.\ (2011), seven stars have an effective temperature above
20\,000\,K and are positioned in the joint part of the $\beta\,$Cep and SPB
instability strips, but only one of these seven has a temperature close to the
one of HD\,180642. This star's variability is probably due to rotation and at
the mmag level, i.e., a factor of more than ten below the one of our target, so
the comparison between the Kepler stars and HD\,180642 in terms of solar-like
oscillations is premature.

On the observational side, we made an extensive search for ridges in
  \'echelle diagrams constructed for the CoRoT residual light curve of
  HD\,180642 discussed in Paper\,I, adopting the same method as in the case of
  the CoRoT data of the O8.5V pulsator HD\,46149 (Degroote et al.\ 2010b).  In
  doing so, we considered numerous values for a large frequency
  spacing. We failed to identify any ridges, in particular also for the value of
  the spacing reported in Belkacem et al.\ (2009).  This model-independent
  result stands in contrast to the clear ridges present in the \'echelle diagram
  of HD\,46149.  A comparison between the \'echelle diagram of HD\,46149 and two
  for HD\,180642, one based on the large spacing by Belkacem et al.\ (2009) and
  the most likely one reported in Paper\,I, is shown in Fig.\,\ref{echelle}.
  The recently discovered solar-like oscillations in the frequency regime above
  that of the heat-driven modes in a $\delta\,$Sct star observed with the
  {\it Kepler\/} satellite did lead to clear ridges in the \'echelle diagram of
  that star (Antochi et al.\ 2011), as is the case for HD\,46149.  

We then considered our seismic models that explain all the modes of HD\,180642
in the frequency regime $[0,20]\,$\cd in terms of heat-driven modes with
non-linear mode couplings (listed in Table\,\ref{models}).  Irrespective of the
cause of the power in the range $[11,26]\,$\cd, we first computed all frequency
differences between the axisymmetric mode frequencies of consecutive radial
order for the same degree $\ell$ in that interval and compared all these spacing
values with the one reported by Belkacem et al.\ (2009), with an allowed
uncertainty of 0.02\,\cd.  For the model with $M=11.4\,$M$_\odot$, this spacing
occurs only once, i.e., between the $\ell=3,m=0$, g$_1$ and g$_2$ modes. It also
occurs once between the $\ell=3,m=0$, p$_3$ and p$_4$ modes of the model with
$M=11.8\,$M$_\odot$, while it never occurs for the three other models in the
upper part of Table\,\ref{models}.  Subsequently, we computed all frequency
differences for all the axisymmetric modes of $\ell=0,\ldots,4$ in the frequency
interval $[11,26]\,$\cd, to test whether any particular spacing occurred.  The
outcome of these computations is shown in Fig.\,\ref{spacingfig} for the two
models indicated in italics in Table\,\ref{models}. It can be 
seen that no
frequency spacing can be discerned among the model frequencies.

\section{Concluding remarks}

We have managed to explain the extensive observational seismic input deduced from
CoRoT and ground-based photometry along with high-resolution spectroscopy of the
$\beta\,$Cep star HD\,180642 by means of present-day models. All of the
30 detected frequencies are fitted by model frequencies, of which 16 are
predicted to be excited and 13 are combination frequencies that can be linked
to the dominant radial fundamental mode with an extremely large amplitude
for this type of pulsator.  Our modelling resulted in a mass between
11.4 and 11.8\,M$_\odot$ and a zero or very low core overshooting parameter,
along with an age of 
between 12 and 13 million years (corresponding to a central
hydrogen fraction between 
0.21 and 0.25) if we allow an excitation problem for only one
of the independent oscillation modes, namely for the $\ell=3$, p$_1$ mode with a
frequency 8.77\,\cd. This is a frequency regime where similar excitation
problems have been found for this type of pulsator (e.g., Dziembowski \&
Pamyatnykh 2008, Zdravkov \& Pamyatnykh 2009).  All other observed frequencies
that are not excited by the heat mechanism can be explained as eigenmodes 
triggered by the large-amplitude radial mode by means of non-linear resonant mode
excitation.

We are unable to support the findings of Belkacem et al.\ (2009) from
the seismic models which explain the oscillation spectrum of HD\,180642 in the
frequency regime $[0,20]\,$\cd, i.e., the nature of the power excess in the
p-mode frequency regime of HD\,180642 is different from the one of ``classical''
solar-like oscillations caused by the 
stochastic forcing due to convective motions.
We instead attribute the occurrence of time-dependent amplitudes and
phases in the frequency regime above the radial fundamental mode to the effect
of the non-linear resonant mode coupling which is activated by the large
amplitude of the dominant mode. This provides a natural explanation for the lack
of detections of solar-like oscillations in B stars monitored by both the CoRoT
and {\it Kepler\/} missions. All of the B pulsators observed by these
satellites indeed 
have much lower amplitudes (at least by a factor ten below the dominant
mode of HD\,180642) so resonant mode coupling is not expected for these, while
solar-like oscillations would have been detected if they existed in such types
of stars.

We find, once more, a discrepancy between the spectroscopic $\log\,g$ and the
seismic $\log\,g$, as was found before in two other seismically modelled OB
stars. We assign this to the systematic uncertainty when deducing $\log\,g$ from
the wings of spectral lines of one snapshot spectrum or from an average of a
time series of spectra. In both cases, the pulsational broadening is not taken
into account in an appropriate way. Indeed, oscillations change the shape of
spectral lines in an asymmetric and time-dependent way, i.e., the level of
deformation depends on the phase in the overall oscillation cycle. To compensate
for this deformation and absence of pulsational broadening, one usually
introduces a time-independent ad-hoc parameter called macroturbulence (e.g.,
Aerts et al.\ 2009) and/or a rotational profile with a wrong $v\sin\,i$, each of
which are usually deduced from metal lines and have a different functional shape
than the true time-dependent line-broadening function. This is then compensated
for by an incorrect value of the gravity and/or microturbulence, which can
either be too high or too low.  In addition to this, the normalisation of the
spectra, which is a necessary manipulation of the data required to deduce the
position and shape of the line wings with respect to the continuum, is by no
means a trivial operation and in itself introduces systematic uncertainty.  In
the present case of HD\,180642, for instance, the gravity was deduced from four
Balmer line wings assuming a rotational profile of 44\,km\,s$^{-1}$, while a
problem was encountered in deriving the microturbulence in a consistent way,
because a discrepancy of 5\,km\,s$^{-1}$ occurred between the values deduced
from the N and O lines for this quantity.  Our seismic analysis inferred that
$v\sin\,i=25\,$km\,s$^{-1}$ only. There are thus various uncertainties to be
considered for quantities that are not independent. The seismic gravity is a
quantity determined by the entire stellar structure, which is very strongly
constrained for HD\,180642 by the numerous seismic properties measured for this
star.  In view of the discrepancy, we advise the use spectroscopically derived
$log\,g$-values of pulsators with a $3\sigma$ error bar. Temperature estimates
are not affected by this uncertainty as they rely on an ionisation balance of Si
for which three stages are available in the case of $\beta\,$Cep stars, hence
this quantity is not very dependent on the line shapes.

\begin{acknowledgements}
  The research leading to these results has received funding from the European
  Research Council under the European Community's Seventh Framework Programme
  (FP7/2007--2013)/ERC grant agreement n$^\circ$227224 (PROSPERITY), from the
  Research Council of K.U.Leuven grant agreement GOA/2008/04 and from the
  Belgian Federal Space Office under contract C90309: CoRoT Data Exploitation.
  M.B.\ and C.A.\ acknowledge the Fund for Scientific Research -- Flanders for a
  grant for a long stay abroad and for a sabbatical leave, respectively.
\end{acknowledgements}

{}

\end{document}